\documentclass[conference]{IEEEtran}

\usepackage{bm}
\usepackage{wrapfig}
\usepackage{float}
\usepackage{amsmath}
\usepackage{amssymb}
\usepackage{multirow}
\usepackage{amsfonts}
\usepackage{mathrsfs}
\usepackage{graphicx}
\usepackage{epstopdf}
\usepackage{caption}
\usepackage{subcaption}
\usepackage{adjustbox}
\usepackage{textcomp}
\usepackage{autobreak}
\usepackage{tabularx}
\usepackage{comment}
\usepackage{lettrine}
\usepackage{cite}
\usepackage{setspace}
\usepackage{inputenc}
\usepackage{fancyhdr}
\usepackage{afterpage}
\ifCLASSINFOpdf
\else
\fi

\fancypagestyle{firststyle}
{
   \pagestyle{fancy}
   \fancyfoot{}
\lhead{\sc{Presented at the 11Th Bulk Power Systems Dynamics And Control Symposium (IREP 2022), July 25-30, 2022, Banff, Canada}}
   
}

\begin{document}
%

\title{Influence of Stochastic Dependence on Network Constraints Screening for Unit Commitment}

\author{
\IEEEauthorblockN{Mohamed Awadalla and François Bouffard}
\IEEEauthorblockA{Department of Electrical and Computer Engineering\\
McGill University\\
Montreal, QC, H3A 0E9, Canada\\
Email: mohamed.awadalla@mail.mcgill.ca, francois.bouffard@mcgill.ca}
}

%

\maketitle

\thispagestyle{firststyle}
\begin{abstract}

The deepening penetration of renewable energy is challenging how power system operators cope with the associated variability and uncertainty in the unit commitment problem. Given its computational complexity, several optimization-based methods have been proposed to lighten the full unit commitment formulation by removing redundant line flow constraints. These approaches often ignore the spatial couplings of multi-side renewable generation and demand. To address this pitfall, we rule out redundant constraints over a tightened linear programming relaxation of the original unit commitment feasibility region by adding a constraint that efficiently models the correlation of residual demand variations. We set forth a novel, tractable and robust polyhedral uncertainty envelope induced by a given set of scenarios to characterize the tightening constraint. We propose a data-driven umbrella constraint discovery problem formulation that  substantially increase the network constraints filtration in unit commitment. Numerical tests are performed on standard IEEE test networks to substantiate the effectiveness of the approach.

\end{abstract}

%
\IEEEpeerreviewmaketitle

\section*{Nomenclature}

The main symbols used in the paper are listed here. Further symbols will be defined as required.

\subsection{Sets}
\begin{IEEEdescription}[\IEEEusemathlabelsep\IEEEsetlabelwidth{$1,2,3,4$}]
\item[$\mathcal{N}$]{Set of buses, indexed by $n$.}
\item[$\mathcal{L}$]{Set of transmission lines, indexed by $l$.}
\item[$\mathcal{M}$]{Set of generating units, indexed by $m$.}
\item[$\mathcal{M}_{n}$]{Set of generating units connected to the node $n$.}
\item[$\mathcal{T}$]{ Set of time periods, indexed by $t$.}
\end{IEEEdescription}

\subsection{Parameters}
\begin{IEEEdescription}[\IEEEusemathlabelsep\IEEEsetlabelwidth{$1,2,3,4$}]
\item[$c_{m}$]{Production cost of the generating unit $m$.}
\item[$d_{n}$]{Residual demand at node $n$.}
\item[$f_l^{max}$]{Maximum flow capacity of transmission line $l$.} 
\item[$g_m^{max}$]{Maximum power limit of generator $m$.}
\item[$g_m^{min}$]{Minimum power limit of generator $m$.}
\item[$h_{l n}$]{Power transfer distribution factor (PTDF) of line $l$ at bus $n$.}
\end{IEEEdescription}

\subsection{Variables}
\begin{IEEEdescription}[\IEEEusemathlabelsep\IEEEsetlabelwidth{$1,2,3,4$}]
\item[${g}_{m}$]{Power output dispatch of generating unit $m$.}
\item[${q}_{n}$]{Net injected power at node $n$.}
\item[${u}_{m}$]{Commitment of generating unit $m$.}
\end{IEEEdescription}

\section{Introduction}

Despite the liberalization of the electricity sector, unit-commitment (UC) is still a fundamental optimization tool for all major independent system operators in the US, such as ISO New England, CAISO and PJM, for the daily operation of power systems \cite{Sun2017, Zheng}. The solution to the UC problem determines the most economical operating schedule, given by the on/off commitment status and production levels of the generating units. The goal of the UC problem is to minimize system operational cost, while satisfying generation, network and various reliability and security constraints. \par

Mathematically, UC is generally formulated as a mixed-integer programming problem (MIP), which belongs to the class of NP-hard problems \cite{anjos}. Moreover, penetration of renewable energy in power systems is rapidly increasing.
Such a growth of solar and wind power, with their stochastic characteristics, poses uncertainty to the conventional UC formulation. Therefore, developing computationally efficient and robust methods to solve the UC problem to optimality has been and continues to be a hot research topic. \par 

Power system operators experience have shown that in the UC problem only a very small proportion of network constraints are indeed congested \cite{Ardakani}. Considering only the potentially active transmission constraints significantly reduces the UC solution times and  can facilitate obtaining the optimal UC plan. In the literature, several optimization-based methods have been proposed for constraints screening for the UC problem. Constraint generation is a well-known iterative procedure where the violated constraints from the original UC are gradually added to the reduced one until the solution to the latter is feasible in the former \cite{Fu}. This method is applied to security constrained unit commitment in \cite{Tejada-Arango} to filter out post-contingency constraints. The drawback of the constraint generation method is that, it is computationally expensive if the required number of iterations to guarantee a feasible solution is large. The notion of umbrella constraints has been introduced for identifying redundant constraints which do not alter the feasibility region of the original UC problem \cite{Ardakani}. Similarly, references \cite{Zhai, Roald} solve two optimization problems for each line to remove as many redundant constraints as possible from UC formulation. \par

In addition, the variability and uncertainty of renewable power generation have introduced new challenges into the UC problem \cite{Mohandes}. With a large-scale uptake of uncertain generation, advancement in two aspects is required to close the gap between computationally tractable and robust solutions to the UC constraints screening problem. First, it is necessary to design a computationally tractable scalable optimization approach that can handle a high degree of uncertainty. Second, it is critical to create a robust representation of the uncertainty set that can be used as an input to the aforementioned optimization strategy \cite{Golestaneh}. Previous work \cite{Zhai, Ardakani} developed a network screening algorithm with a single nodal (residual) demand vector, while references \cite{Roald,Amir2} considered residual demand as a decision variable in the optimization problem. However, these studies built box shaped uncertainty set for the univariate system net load (i.e., load less renewable generation). \par

In light of this shortcoming, we extend the notion of the umbrella constraint discovery formulation of Abiri-Jahromi and Bouffard \cite{Amir2}, by comprehensively capturing the spatial correlation of geographically distributed renewable generation and loads in a power system. In this spirit, we present a novel modelling technique for uncertainty sets which we call \emph {data-driven polyhedral uncertainty envelope}. Compared to previous work, our proposed uncertainty set is less conservative than the box shaped envelope of \cite{Ardakani, Roald} and computationally cheaper than computing the convex hull of the uncertainty data \cite{Golestaneh,Velloso}. \newline

\section{Unit Commitment Model}\label{FE_transmission}
\subsection{Introduction}
For expository purposes, we use a simplified single-period unit commitment optimization framework \cite{Pineda1} according to the following simplifying assumptions:\par

\begin{itemize}
\item Single-period: UC is usually formulated as a multi-period problem that incorporates inter-temporal constraints. However, since this work focuses on investigating the impact of residual demand spatial correlation on network reduction, we prefer to investigate such an effect solely by considering a single-period UC as in \cite{Pineda1, Roald, Amir2}. 
\item DC power flow: The power flows in the transmission  lines are estimated via a DC approximation by using power transfer distribution factors (PTDF) to keep the model linear. The PTDF of line $l$ with respect to node $n$ is
denoted as $h_{l n}$. Besides, $f_{l}^{max}$ represents maximum flow capacity of transmission line $l$. The number of buses and lines are denoted by $\mathcal{N}$ and $\mathcal{L}$, respectively. 

\item Generation portfolio: Each generating unit $m$ is characterized by a minimum and a maximum power output which denoted as $g_{m}^{min}$ and $g_{m}^{max}$, respectively. 
\item Residual demand: The demand vector $d_n$ is modelled as a net load (load less renewable generation). Without loss of generality, we assume that historical net-load are multivariate normally distributed scenarios.     
\item No contingencies: We assume that all generators and lines are fully operational and therefore security constraints are neglected.\newline
\end{itemize}

\subsection{Problem Formulation}
The optimization problem corresponding to this simplified UC is a Mixed Integer Linear Programming (MILP) problem and  formulated as \cite{Pineda1}:


\begin{equation}
\min _{u_{m}, g_{m}, q_{n}} \sum_{m \in \mathcal{M}} c_{m} g_{m}
\end{equation}
Subject to:

\begin{equation}
 \quad q_{n} =\sum_{m \in \mathcal{M}_{n}} g_{m}-d_{n} \quad \forall n \in\mathcal{N}
\end{equation}

\begin{equation}
\sum_{n \in \mathcal{N}} q_{n}=0
\end{equation}

\begin{equation}
u_{m} g_{m}^{min} \leq g_{m} \leq u_{g} g_{m}^{max} \quad \forall m \in\mathcal{M}
\end{equation}

\begin{equation}
-f_{l}^{\max } \leq \sum_{n \in \mathcal{N}} h_{l n} q_{n} \leq f_{l}^{\max } \quad \forall l \in\mathcal{L}
\end{equation}

\begin{equation}
u_{m} \in\{0,1\} \quad \forall m \in\mathcal{M}
\end{equation}

Decision variables include the commitment status of the generating units $u_{m}$, the power output schedules  $g_{m}$, the net power injections at each node $q_{n}$. The objective function (7) minimizes the total generation cost. Constraint (8) computes the net injected power at each node, while constraint (9) ensures power balance in the system. Constraints (10) and (11) enforce limits on generators outputs and power flows on transmission lines using PTDF, respectively. Finally, (12) defines the binary variables of the generating units.\newline

\section{Data-Driven Residual Demand Uncertainty Set Characterization}

The procedure for building a data-driven polyhedral uncertainty envelope is introduced next. Without loss of generality, we focus on capturing the spatially correlated uncertain residual demand. We develop a scenario-based polyhedral uncertainty set \cite{Velloso} by leveraging Principal Component Analysis (PCA) \cite{Huo, Berizzi}. PCA applies to time series of uncertain load demand that contains their current and historical data. All the input time series have the same length of T , and they are synchronized and evenly spaced in time (e.g. one hour
interval).  \par

We denote $W$ as a matrix whose elements $w_{n t}$ are the input time series of residual demand power at bus $n \in \{1,...,\mathcal{N}\}$ for each time instance $t \in \{1,...,\mathcal{T}\}$, ${W} \in {\mathbb{R}}^{\mathcal{N}\times\mathcal{T}}$. We denote $M$ as the number of available historical scenarios which are embedding relevant information about the true underlying correlation between uncertain parameters. We compute the sample mean and centered input data matrix $W_{c}$ as follows \cite{Huo, Berizzi}: 


\begin{equation}
\mu_{n}=\frac{1}{M} \sum_{t=1}^{M} w_{n t} 
\end{equation}

\begin{equation}
W_{c}=W-( \mu \cdot 1^\top)
\end{equation}
where $\mu$ is a column vector of the sample mean at each bus $n$, $\mathbf{1}$ is a $\mathcal{N} \times 1$ column vector of ones. Then, the covariance matrix $\Sigma \in {\mathbb{R}}^{\mathcal{N}\times\mathcal{N}}$ is approximated using $M$ scenarios as follow:

\begin{equation}
\Sigma=\frac{1}{M-1} {W}_{c}^{\top} {W}_{c}
\end{equation}

The PCA is performed by conducting the eigenvalue decomposition on the covariance matrix. Letting the columns of $V$ and the diagonal entries of $\Lambda$ represent the eigenvectors and eigenvalues of $\Sigma$, respectively. The dimensions of $\Lambda$ and $V$ matrices is $\mathcal{N} \times \mathcal{N}$. The principal component coefficients of the covariance matrix can be derived as \cite{Berizzi}:

\begin{equation}
{Z}= {V}^{\top}{W}_{c}
\end{equation}

where $Z$ is principal components matrix with size $\mathcal{N}\times \mathcal{T}$, the $n$th row of matrix $Z$, $z_{n}$, is the $n$th PC. The reader can refer to \cite{Berizzi} for full derivation of the PC computation technique. The reconstruction of net demand data from only first $\mathcal{K}$ $(\mathcal{K} < \mathcal{N})$ components is implemented inversely as \cite{Berizzi}:

\begin{equation}
\hat{W}_{k}={V}_{k}Z_{k} + ( \mu_{n}\cdot 1)
\end{equation}

where $\left\{\hat{W}_{k}\right\}_{k=1, \ldots, \mathcal{K}}$ is the set of net demand data reconstruction corresponding to each principal component and eigenvector $k$, which is hereinafter referred to as scenarios $\mathcal{S}$. The number of $\mathcal{S}$ scenarios is denoted by $\mathcal{I}$ and directly proportional to the choice of principal components $\mathcal{K}$. Inspired by the data-driven convex hull uncertainty set \cite{Velloso,Golestaneh}, and by applying PCA to historical scenarios according to the previous steps, we propose a scenario-induced PCA-based \emph{polyhedral uncertainty envelope (PUE)} as follows:


\begin{equation}
\begin{gathered}
{\mathbf{E}}(\mathcal{S})= \left\{\mathbf{E} : \mathbf{E}=\sum_{i=1}^{\mathcal{I}} \lambda_{i} \mathcal{S}_{i}, \sum_{i=1}^{\mathcal{I}} \lambda_{i}=1, \lambda_{i} \geq 0\right\}
\end{gathered}
\end{equation}

\begin{figure}[t!]
\includegraphics[width=8.8cm]{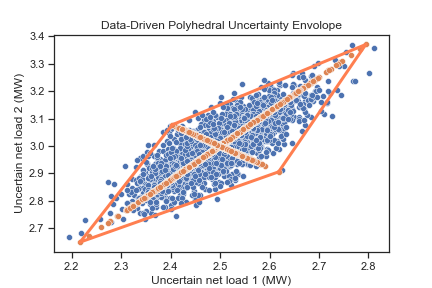}
\caption{Schematic illustration of polyhedral uncertainty envelope.}
\label{Fig_system}
\end{figure}

Note that \emph{PUE} is a convex hull of $\mathcal{S}$, thereby representing the smallest convex set that contains every scenario in $\mathcal{S}$. Fig. 1 illustrates the construction of \emph{PUE} in two dimensions. In this example, the original data and reconstructed scenarios from two principal components indicated by blue and orange dots, respectively. By inspection, the envelope encapsulates the majority of the correlated scenarios and captures accurately the largest and smallest projected scenarios onto each principal component direction. This proposed uncertainty model has two main advantages compared to the convex hull \cite{Velloso}. First, the number of edges depends on the chosen $\mathcal{K}$ principal components and maximum can be 2$\mathcal{N}$. Conversely, the convex hull is sensitive to all possible scenarios that a power system might encounter, albeit unlikely. This makes the convex hull clearly affected by the outliers, and consequently the number of vertices can grow exponentially when the uncertainty dimensionality increases. Second, through selecting certain number of PCs components, \emph{PUE} allows for direct trade-offs between tractability and conservativeness when compared to convex hull. \newline

\section{Umbrella Constraint Discovery}
\subsection{Identification of Umbrella Network Constraints}

The UC problem (1)-(6) can be made significantly easier to solve if constraints (5) with no impact on the optimal UC plan are removed. In this paper, we tailor the notion of Umbrella Constraint Discovery (UCD) for the purpose of identifying potentially active network constraints in unit commitment. \par

UCD is an iterative approach to identify the set of umbrella line constraints that minimally characterize the feasible region of the original problem (1)-(6) \cite{Amir2}. At each iteration, an optimization problem is solved to pinpoint the set of line constraints forming one of the vertices of the feasible region of the original UC problem. For computational tractability, UCD is solved using the LP-relaxation of the feasible region of the UC problem instead of the original one. Particularly, at each iteration, we solve for the binary vectors $v_{l} \in\{0,1\}^{\mathcal{L}}$, $u_{g} \in\{0,1\}^{\mathcal{G}}$ and continuous vectors $q,d \in \mathbb{R}^{\mathcal{N}}$, $g \in \mathbb{R}^{\mathcal{M}}$ and $z \in \mathbb{R}_{+}^{\mathcal{L}}$, where $\Omega$ is a large positive number. \par

\begin{equation}
\min _{u_{m}, g_{m}, q_{n}, d_{n} ,v_{l}, z_{l}} \sum_{l \in \mathcal{L}} v_{l}
\end{equation}

Subject to:

\begin{equation}
(2)-(4)
\end{equation}

\begin{equation}
-f_{l}^{\max } \leq \sum_{n \in \mathcal{N}} h_{l n} q_{n} \leq f_{l}^{\max } \quad \forall l \in\mathcal{L}
\end{equation}

\begin{equation}
-f_{l}^{\max } \geq \sum_{n \in \mathcal{N}} h_{l n} q_{n} + z_l \geq f_{l}^{\max } \quad \forall l \in\mathcal{L}
\end{equation}

\begin{equation}
v_{l} - \frac{z_{l}}{\Omega} \geq 0 \qquad \forall l \in\mathcal{L}
\end{equation}

\begin{equation}
z_{l} \geq 0 \qquad \forall l \in\mathcal{L}
\end{equation}

\begin{equation}
v_{l} \in\{0,1\} \qquad \forall l \in\mathcal{L}
\end{equation}

\begin{equation}
0 \leq u_{m} \leq 1 \quad \forall m \in \mathcal{M}
\end{equation}

\begin{equation}
d_{n}^{\min } \leq d_{n} \leq d_{n}^{\max } \qquad \forall n \in\mathcal{N}
\end{equation}

To solve (13)–(21), we apply the iterative procedure described in \cite{Amir2} to identify potentially active line flow constraints considering the generation and demand limits. \par

The objective function (13) aims to minimize the sum of the binary variables, $v_{l}$, by finding the maximum number of line flow constraints that can hit or surpass line capacity. The set of constraints (2)-(4) from the  UC, which controls the decision variables $q_{n}$ and $g_{m}$ has also been forced in this problem. Each line flow constraint (15) is combined with a constraint (16). By inspection, the auxiliary variable $z_{l}$ can be equal to zero only if there is a $q$ satisfying both (15) and (16). Hence, the binary variable $v_{l}$ = 0. Contrarily, $z_{l}$ has to be positive and the binary variable $v_{l}$ = 1 as modelled in (17). Furthermore, the constraints for which $v_{l}$ = 1 have to intersect at the same value of $q$. Since $q$ is an intersection of line constraints over LP-relaxation of the feasible set of the UC problem, it is therefore a vertex of this set. \par

Following the first iteration, which revealed the vertex with the most intersecting constraints, the next step is to pinpoint the other vertices that have the same or less intersecting umbrella constraints. This is done while the previously discovered umbrella constraints are removed from the search by setting their binary variable equal to 1. We terminate the search when there is no more umbrella constraints to identify. Additionally, the vector of residual demand at each bus $n$ is turned into a vector of decision variables as in (21). \par
 
The drawback of UCD relates to the modeling of residual demand range limits in constraint (21). For instance, redundant network constraints have been filtered out considering univariate system net load \cite{Amir2}. This result will be sub-optimal because it ignores the spatial correlation among the residual demands. Therefore, we propose  Data-Driven UCD (DD-UCD) where residual demand vector is characterized using polyhedral uncertainty set $d_{n} \in PUE$ as described in Section III, rather than box shape uncertainty set in (21). \newline

\subsection{Benchmark Method (BN): Roald's Method}
This method has been proposed in \cite{Roald}, it is based upon the solution of one maximization and one minimization for each transmission line $\hat{l}$, these two optimizations are jointly formulated as 

\begin{equation}
\max _{u_{g}, p_{g}, d_{n}, q_{n}} / \min _{u_{g}, p_{g}, d_{n}, q_{n}} \sum_{n \in \mathcal{N}} a_{\hat{l} n} q_{n}
\end{equation}

\begin{equation}
(2)-(4)
\end{equation}

\begin{equation}
-f_{l}^{\max } \leq \sum_{n \in \mathcal{N}} h_{l n} q_{n} \leq f_{l}^{\max } \quad \forall l \in\mathcal{L}, l \neq \hat{l}
\end{equation}

\begin{equation}
0 \leq u_{m} \leq 1 \quad \forall m \in \mathcal{M}
\end{equation}

\begin{equation}
d_{n}^{\min } \leq d_{n} \leq d_{n}^{\max } \qquad \forall n \in\mathcal{N}
\end{equation}

Problem (22)-(26) seeks to maximize/minimize the power flow through each transmission line $\hat{l}$ over an LP-relaxation of the feasible region of the UC problem. \newline

\section{Case Study}
\subsection{Performance Evaluation}
The procedure to measure the performance of the approaches
described in Section IV to rule out line flow constraints is described as follows:

\begin{enumerate}
 \item Given the historical data, the set of retained network constraints from each approach is denoted by $\mathcal{L}^{a}$, and the percentage of umbrella line flow constraints denoted by $\mathcal{R}^{a}=100 \cdot \mathcal{L}^{a} / \mathcal{L}$.
 \item The computational time needed to screen out the redundant network constraints using each screening approach. \newline
\end{enumerate}

\subsection{Simulation Setup}
The DD-UCD and BN screening approaches were formulated as MILP problems. The solution optimality gap was set to 0\% for the two approaches. The simulations have been performed on GAMS platform using CPLEX. The computer used is equipped with an Intel Core i7 3.10 GHz processor and 16 GB RAM.\newline

\subsection{Numerical Results}
In this section, we provide simulation results from four standard IEEE test networks, namely the IEEE-RTS-24, IEEE-57, IEEE-RTS-73, and IEEE-118 test systems \cite{Zimmerman}. Moreover, we assess the computational complexity of each screening approach. All the technical data related to these systems are available in \cite{data} and their main features are listed in Table 1. For the IEEE-RTS-24 and the IEEE-57, the line capacities are reduced by half to have a more congested network. \par

\begin{table}[t!]
\captionsetup{font=scriptsize}
\caption{DESCRIPTION OF TEST POWER SYSTEMS} 
\centering 
\scalebox{0.8}{
\begin{tabular}{c c c c} 
\hline
System & \# Nodes & \# Generators & \# Lines \\ [0.5ex] 
\hline 
IEEE-RTS-24     & 24   & 32 & 38  \\ [0.5ex]
IEEE-57         & 57   & 4  & 41  \\ [0.5ex]
IEEE-RTS-73     & 73   & 96 & 120 \\ [0.5ex]
IEEE-118        & 118  & 19 & 186  \\ [0.5ex] 
\hline 
\end{tabular}}
\label{table:nonlin} 
\end{table}

\begin{table}[t!]
\captionsetup{font=scriptsize}
\caption{SOLUTION TIMES FOR CONSTRAINTS SCREENING APPROACHES} 
\centering 
\scalebox{0.8}{
\begin{tabular}{c c c} 
\hline
\multicolumn{3}{c}{Screening time (mm:ss.ss)} \\
\hline 
Method & DD-UCD & BN   \\ [0.5ex] 
\hline 
IEEE-RTS-24 & 00:4.538   & 1:12.048 (1487\%)  \\ [1ex]
IEEE-57      & 00:4.02   & 3:39.21 (4975\%)  \\ [1ex]
IEEE-RTS-73  & 00:49.22  & 5:54.13  (575\%)   \\ [1ex]
IEEE-118     & 01:4.42   & 8:46.073 (688\%)   \\ [1ex] 
\hline 
\end{tabular}}
\label{table:nonlin} 
\end{table}

\begin{table*}[t!]
\captionsetup{font=scriptsize}
\caption{NUMBER OF UMBRELLA LINE FLOW CONSTRAINTS IDENTIFIED FROM EACH SCREENING APPROACH} 
\centering 
\scalebox{0.8}{
\begin{tabular}{ccccccc} 
\hline
\multicolumn{6}{c}{$Percentage$ $of$ $Umbrella$ $Constraints$ $R_{a}$($\%$)} \\
\hline
Method & DD-UCD & BN & DD-UCD & BN & DD-UCD & BN \\ [1.0ex] 
\hline 
Uncertainty (\% of $\mu$) & \multicolumn{2}{c}{Congested IEEE 24-RTS} & \multicolumn{2}{c}{IEEE RTS-96} & \multicolumn{2}{c}{IEEE 118
Bus} \\ [1.0ex] 
\hline  
10\% & 9.21   & 9.21  (0.0\%)     & 2.08  & 5.83  (+180.0\%)  & 7.12  & 8.06 (+13.2\%)  \\[1ex]
20\% & 11.84  & 21.05 (+77.78\%)  & 5     & 14.16 (+183.3\%)  & 7.25  & 9.4 (+29.62\%)  \\[1ex]
30\% & 15.78  & 26.31 (+66.67\%)  & 7.08  & 19.16 (+170.58\%) & 7.93  & 10.88 (+37.28\%) \\[1ex]
40\% & 17.10  & 38.15 (+123.07\%) & 9.16  & 24.58 (+168.18\%) & 8.06  & 12.5 (+55\%) \\[1ex]
50\% & 19.73  & 42.10 (+113.33\%) & 10    & 27.08 (+170.83\%) & 8.87  & 13.7 (+54.54\%) \\[1ex]
\hline 
\end{tabular}}
\label{table:nonlin} 
\end{table*}

Finally, we synthetically generate correlated, normally distributed residual demands scenarios for 2000 time instances as follows. First, we generate uncorrelated scenarios using $N\left(\mu_{n}, \sigma_{n}^{2}\right)$, $ \forall n \in\mathcal{N}$. The mean residual demand $\mu_{n}$ for each test system are set at the nominal demand values provided in \cite{Zimmerman}. The standard deviation vector, $\sigma_{n}$, is approximated from the mean values using the 3$\sigma$ empirical rule \cite{Qi}. Second, the correlation between any two residual demands is estimated in the covariance matrix by multiplying their independent standard deviations with a weighting factor $\rho = 0.8$. Different levels of uncertainty ranging from 10\% to 50\% of the residual demand mean are adopted for modelling load variations. We have utilised all the principal components to construct the polyhedral uncertainty envelop in DD-UCD approach. 

First, one of the critical advantages of the proposed approach is that in terms of run time, it is significantly faster than Roald's method, as seen in Table II. The run times for DD-UCD approach varied from 4.02–64.42 seconds, while for Roalds's approach from 72.048–526 seconds. Therefore, the proposed approach for the stressed IEEE-RTS-24 and the IEEE-118 buses is 1487\% and 688\% faster compared to Roald's method, respectively. While for IEEE-RTS-73, the proposed approach is 575\% faster compared to the BN method. The DD-UCD identifies all the umbrella line flow limits of the IEEE-RTS-24 and IEEE-118 bus test systems in two and nine iterations, respectively. On the other hand, Roald's (BN method) optimization problem solves optimization problem (22)-(26) $2 \mathcal{L}$ times for each test network. Hence, it requires $76$ and $372$ iterations for the IEEE-RTS-24 and the IEEE-118 bus test system, respectively. \par 

The main outcome of Table II is that the UCD algorithm can significantly accelerate the screening time of redundant constraints compared to the state-of-art approach. Also, Table II clearly shows that the solution time of BN is directly proportional to the number of lines while our proposed approach is simply not. UCD screening time is mainly affected by the final number of discovered umbrella network constraints and the system size. Generally, when the network size grows, it's expected to have more redundant constraints and fewer number of umbrella network constraints \cite{Ardakani}. Thus, the UCD approach is potentially more scalable when applied to practical power systems. Finally, UCD is recursive algorithm and the full assessment of computational complexity in practical networks is part of our ongoing research work. \par

For the identification of umbrella network constraints, DD-UCD and Roald's method are solved for the congested IEEE-24-RTS, IEEE-RTS-73 and IEEE-118, respectively. Mainly, we investigate the impact of a range of residual demand from $\pm 10\%$ to $\pm 50\%$ variation around the nominal values specified in MATPOWER datasets \cite{Zimmerman}. For all test networks, we noticed an increasing trend in umbrella line flow constraints results from both approaches when we increase the level of uncertainty. This is a direct result as we cover wider residual demand ranges where the possibility of hitting line flow limit is higher. Nevertheless, Our proposed approach significantly outperforms Roald's method in the percentage of network constraints removal considering different uncertainty levels. Our proposed approach yields only 10\% and 8.8\% of the total number of constraints compared to 27\% and 13.7\% while considering highest level of uncertainty for IEEE-RTS-73 and IEEE-118, respectively. In congested IEEE-RTS-24, we achieved the highest umbrella network constraints discovery, 19.7\%, considering maximum uncertainty range, while the benchmark method shows 42.10\% with an increase approximated by 113\% compared to our proposed approach. From Table III, DD-UCD shows a reduction in the number of umbrella constraints compared to the benchmark approach by excluding the lines that will not be congested based on the given set of data. Also, the results shown in Table III emphasizes that our proposed uncertainty envelope is less conservative compared to the box shaped uncertainty set and will save more time later in the UC problem. \newline

Lastly, we consider a test set that consists of 1000 time periods to compare the average computational burden relative to the computational time needed to solve the full UC problem. This computational burden does not include the screening time. The computational time required to run the full UC and reduced UC for IEEE-118 test network is 772.52s and 130.2s, respectively. This significant improvement in computation time will help system operators to run the UC problem much faster compared to current industry practices.     

\section{Conclusion}
Only a small number of transmission line limitations are binding, according to a widespread observation in power system operation and planning optimization. In this paper, we propose a data-driven umbrella constraint discovery approach that takes advantage of historical information to disregard redundant constraints in the UC problem. This approach characterizes polyhedral uncertainty envelope using historical residual demand and principal component analysis. Our approach runs much faster and screen out more network constraints compared to the state-of-art method. \par

Practically, the proposed methodology reduces the UC problem size for system operators, thus improves the solution time of UC problem. Also, DD-UCD is an effective tool in identifying non-umbrella constraints for power system operation and planning problems with high penetration of renewables.  Future work is needed to adapt the proposed approach to the multi-period unit commitment problem, which involves taking into account inter-temporal technical constraints of generating units. It would be interesting to see how the proposed strategy performs in the security-constrained unit commitment problem. Furthermore, the proposed approach may be used to filter out other technical restrictions related to generators unit such as ramping and capacity limits. \newline

\section*{Acknowledgements}
This work was funded in part by IVADO, Montreal, QC and the Natural Sciences and Engineering Research Council of Canada, Ottawa, ON.

\end{document}